\documentclass[prl,twocolumn,english,superscriptaddress,longbibliography]{revtex4}

\usepackage{graphicx}
\usepackage{float}
\usepackage[dvipsnames]{xcolor}
\usepackage{bm,amsmath,amssymb}
\usepackage{soul}
\usepackage[T1]{fontenc}

\usepackage{hyperref}
\hypersetup{
  colorlinks = false,
  urlcolor = black,
}
\usepackage{braket}
\usepackage{xfrac}
\usepackage{verbatim}
\usepackage{lipsum}
\usepackage{cancel}
\usepackage{relsize}
\usepackage{placeins} 
\usepackage{amstext}
\usepackage{bbold}
\usepackage{esint}
\usepackage{tikz}

\newcommand{\kwm}[1]{\textcolor{black}{#1}}

\newcommand{\zww}[1]{\textcolor{black}{#1}}
\newcommand{\rr}[1]{\textcolor{black}{#1}}
\newcommand{\rs}[1]{\textcolor{black}{#1}}

\newcommand{\red}[1]{\textcolor{black}{#1}}
\newcommand{\rv}[1]{\textcolor{black}{#1}}
\newcommand{\rvv}[1]{\textcolor{black}{#1}}

\begin{document}

\title{ Observing a Topological Transition in Weak-Measurement-Induced Geometric Phases}

\author{Yunzhao Wang}
\affiliation{Department of Physics, Washington University, St.\ Louis, Missouri 63130}
\author{Kyrylo Snizhko}
\affiliation{Department of Condensed Matter Physics, Weizmann Institute of Science, Rehovot, 76100 Israel}
\affiliation{Institute for Quantum Materials and Technologies, Karlsruhe Institute of Technology, 76021 Karlsruhe, Germany}
\affiliation{Univ. Grenoble Alpes, CEA, Grenoble INP, IRIG, PHELIQS, 38000}
\author{Alessandro Romito}
\affiliation{Department of Physics, Lancaster University, Lancaster LA1 4YB, United Kingdom}
\author{Yuval Gefen}
\affiliation{Department of Condensed Matter Physics, Weizmann Institute of Science, Rehovot, 76100 Israel}
\author{Kater Murch}
\email{murch@physics.wustl.edu}
\affiliation{Department of Physics, Washington University, St.\ Louis, Missouri 63130}


\date{\today}


\begin{abstract}
 Measurement plays a quintessential role in the control of quantum systems. Beyond initialization and readout which pertain to projective measurements, weak measurements in particular, through their back-action on the system, may enable various levels of coherent control. The latter \rs{ranges} from observing quantum trajectories to state dragging and steering. Furthermore, just like the adiabatic evolution of quantum states that is known to induce the Berry phase, sequential weak measurements \kwm{may lead} to path-dependent geometric phases. Here we measure the geometric phases induced by sequences of weak measurements and demonstrate a topological transition in the geometric phase controlled by measurement strength. This connection between weak measurement induced quantum dynamics and topological transitions reveals subtle topological features in measurement-based manipulation of quantum systems.  \rs{Our protocol could be implemented for classes of operations (e.g.\ braiding) which are topological in nature. Furthermore, our results open new horizons for measurement-enabled quantum control of many-body topological states. } 
\end{abstract}




\maketitle



The geometric phase is a part of the global phase gained by cyclic path of a quantum state, which only depends on the trajectory enclosed by the motion in parameter space \kwm{and not on the traversal time}  \cite{ahar87,cohe19}. \rr{A  frequently} mentioned example is the \kwm{Pancharatnam-Berry}  phase that emerges from adiabatic evolution of the system Hamiltonian \cite{berr84, panc56}. It was suggested that the \kwm{Pancharatnam phase} can be viewed in the framework of strong quantum measurement backaction \cite{berr96}. The latter \cite{katz06,guer07,koro11, jaco14, hatr13, murc13traj} is the inevitable disturbance brought by measurement on a certain quantum system. One example thereof is the projection of a quantum state onto an eigenstate of a strongly-measured observable. \rs{More generally, weak measurements only partially modify the quantum state.  In either case, strong or weak, the accumulated disturbance due to a sequence of measurements can result in closed-path motion of the quantum state.} For a spin 1/2 system the resulting geometric phase is half the solid angle \rr{subtended by the path in parameter space} \cite{facc99,panc56}. \rs{While such measurement-induced geometric phases have been observed in optical systems \cite{berr96,cho19}, a recent theoretical study \cite{gebh19} has pushed this insight to a new qualitative perspective: the emergence of geometrical phases is accompanied by a topological transition \cite{Roushan2014,PhysRevLett.113.050402}.}  In this letter, we utilize a superconducting transmon circuit to \rr{demonstrate and characterize} this measurement-induced topological transition.



\begin{figure}[H]
\centering
\includegraphics[width = 0.5\textwidth]{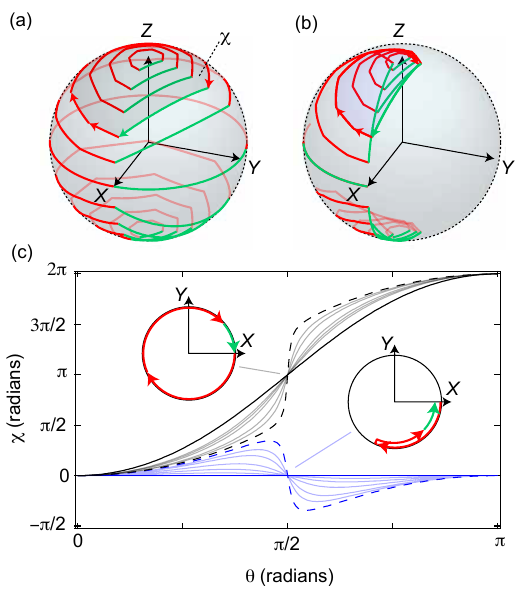}
\caption{\label{fig1}  {\bf Measurement-induced topological transition.} (a) A sequence of  measurements along a fixed latitude drags the state on a trajectory displayed on the \red{surface of the}  Bloch sphere (arrows indicate the backaction of the measurements for the first two, and last (of 6) measurements for one latitude). When an additional, final projective measurement closes the path (green arrow), the state acquires a geometric phase ($\chi$). Considering all latitudes, these trajectories form a closed surface \zww{winding} around the Bloch sphere.  (b) Weaker measurements result in smaller backaction on the state; as a result the trajectories form a closed surface that does not wrap around the Bloch sphere. (c) Dependence of the geometric phase on the polar angle $\theta$ for the measurement sequences with measurement strengths slightly below (blue dashed line) and above (black dashed line) the critical value. The black solid line shows the case of infinitely strong measurements, the blue solid line represents zero measurement strength, and faint lines indicate intermediate measurement strengths. \red{The values of $\chi$ for $\theta = 0$ and $\theta = \pi$ must differ by a multiple of $2\pi$. This difference cannot be changed by continuous deformation of the dependence of $\chi$ on $\theta$. Thus, the behaviors above and below the transition are topologically distinct.}    \kwm{The insets illustrate the origin of the transition. For sufficiently strong measurements the equatorial trajectory circumnavigates the Bloch sphere while for weak measurements  it does not.} }
\end{figure}

We consider a series of variable strength measurements on a pseudo-spin half system. For the spin initialized in a state $|\theta,\phi=0\rangle$, given by polar and azimuthal angles $\theta$ and $\phi$ of the Bloch sphere, a series of measurements along axes with fixed $\theta$ and with $\phi$ ranging from 0 to $-2\pi$ (Fig.~\ref{fig1}a), has the potential to drag the state along the geodesic lines between the axes of consecutive measurements \red{(these geodesic lines, in general, do not retain fixed $\theta$)}.  \kwm{This trajectory results in a geometric phase $\chi$ related to the enclosed solid angle \cite{leek07,fili13,yale16}.  In the limit of continuous strong measurements, $\chi = \frac{1}{2}\times 2\pi(1-\cos\theta)$. However, for weak measurements, the state lags behind the advancing measurement axis and only partially moves along the geodesic line (Fig.~1a).  With an additional, final projective measurement used to close the path, the surface formed by the set of trajectories along different latitudes either ``wraps'' (Fig.~\ref{fig1}a) or does not ``wrap'' the Bloch sphere (Fig.~\ref{fig1}b).}  \red{This property of wrapping or not wrapping cannot be changed by continuous deformations of the set of all trajectories and is linked to the respective topological invariant---the Chern number, which is the equivalent of the winding number for 2\rv{-}dimensional surfaces.  The transition between these two regimes} is controlled by the measurement strength and can be represented \kwm{by a jump in the} Chern number \cite{gebh19}. In Figure 1c, we display the predicted geometric phase $\chi$ versus the polar angle $\theta$ for the extremal cases of strong measurement, zero-strength measurement, and near the topological transition.

\begin{figure}
\begin{center}
\includegraphics[width = 0.5\textwidth]{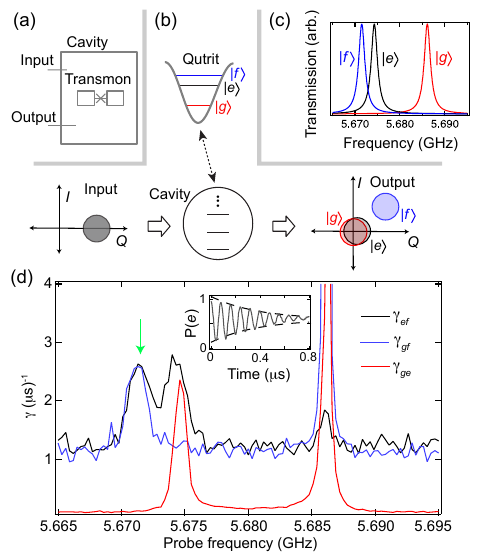}
\end{center}
\caption{\label{fig2}\zww{{\bf Experiment setup.} (a) A superconducting Transmon \rr{qutrit} is dispersively coupled to a high quality factor microwave cavity. (b) \rr{A coherent state probe acquires a qutrit-state ($\ket{g}, \ket{e},$ or $\ket{f}$) dependent displacement on the $I/Q$ plane}.  (c) The dispersive interaction results in a state-dependent cavity transmission, allowing for measurements that resolve one state while leaving the other two \rs{unresolved}. (d) The measurement is quantified via the frequency dependent dephasing rates of \rr{each two-level subspace of} Transmon states. The green arrow indicates the operating probe frequency, which corresponds to a \rr{selective} measurement of the $\ket{f}$ state, which preserves the coherence in the $\{\ket{g},\ket{e}\}$ subspace. A typical Ramsey measurement is shown in the inset. }}
\end{figure}

In order to probe the predicted topological transition, we choose the first three energy levels of a superconducting Transmon circuit \cite{koch07} embedded in 3D aluminium cavity \cite{paik11} as our experiment platform (Fig.~\ref{fig2}a). In the dispersive limit \cite{blai04,wall04}, where the cavity frequency $\omega_\mathrm{r}$ is far detuned from the qutrit transition frequencies $\omega_j$, the Jaynes-Cummings Hamiltonian becomes:
\begin{equation}
\mathrm{H}_m=\hbar\omega_\mathrm{r} a^\dag a+\sum_{j} \hbar \omega_j| j \rangle\langle j|+\sum_{j} \hbar \xi_j |j\rangle \langle j|a^\dag a,
\end{equation}
\rs{where \rs{$a^\dagger a$} is the cavity photon number operator, $\ket{j}$ are the energy eigenstates of the Transmon with energies $\hbar \omega_j$}, and $\hbar \xi_j$ are the interaction energies between the cavity eigenstates and Transmon energy levels $\ket{j}$, and we consider the lowest three energy levels, $j\in\{g,e,f\}$. The effect brought by such interaction energies $\hbar\xi_j$ can be viewed as a qutrit-state-dependent shift on the cavity frequency, enabling quantum non-demolition weak measurements of the circuit energy states. 

As is shown in Figure \ref{fig2}b, when the cavity is probed with a coherent state, the output signal distributes on the quadrature space of the electromagnetic field ($I,Q$) depending on the cavity transmission at the measurement frequency. The setup is operated in the strong dispersive regime, where the cavity linewidth $\kappa \ll \xi_j$. Figure \ref{fig2}c \rs{illustrates that} a weak probe of the cavity near the $\ket{f}$ resonance will be transmitted only if the circuit is in the $\ket{f}$ state. Therefore, the measurement distinguishes the state $\ket{f}$ from both $\ket{e}$ and $\ket{g}$, but does not distinguish $\ket{e}$ from $\ket{g}$. The selective nature of this measurement architecture allows us to reserve one energy level (e.g. $\ket{g}$) as a quantum phase reference in order to determine the global phase accumulated by a state in the $\{\ket{e}, \ket{f}\}$ manifold.  

In the limit $\chi_f\gg \kappa$ the Kraus operators \cite{jaco14} associated with a probe near the $\ket{f}$ resonance are given by
\begin{equation}
M_z^{(\mathbf{r})}= \sqrt{\frac{1}{2 \pi}} \frac{1}{2 \sigma} \begin{pmatrix}  e^{- (\mathbf{r}-\mathbf{r_0})^2/2\sigma^2} & 0 & 0 \\ 0 & e^{-\mathbf{r}^2/2\sigma^2} & 0 \\ 0 & 0 & e^{- \mathbf{r}^2/2\sigma^2} \end{pmatrix}
\end{equation}
in the $\{\ket{f},\ket{e},\ket{g}\}$ energy eigenstate basis. Here $\mathbf{r}$ represents the output signal's location on the $I/Q$ plane, $\mathbf{r_0}$ is the mean output signal when the transmon is in the energy eigenstate $|f\rangle$, $\sigma$ is the variance of the output signal. \rr{A} Kraus operator both gives the probability distribution of a measurement outcome ($P(\mathbf{r}) = |M_z^{(\mathbf{r})} \ket{\psi}|^2$) and characterizes the measurement backaction on the state ($\ket{\psi} \to M_z^{(\mathbf{r})} \ket{\psi}$). \rs{Applying such a measurement pulse of duration $\tau$ reduces coherences between pairs of states, characterized by dephasing factors $\exp(-\gamma_{ef}\tau)= \exp(-\gamma_{gf}\tau)= \exp(-\frac{\mathbf{r_0}^2}{4\sigma^2})$ and $\exp(-\gamma_{ge}\tau)=1$  \cite{Bultink2018}}.
 
 \begin{figure*}
\begin{center}
\includegraphics[width = \textwidth]{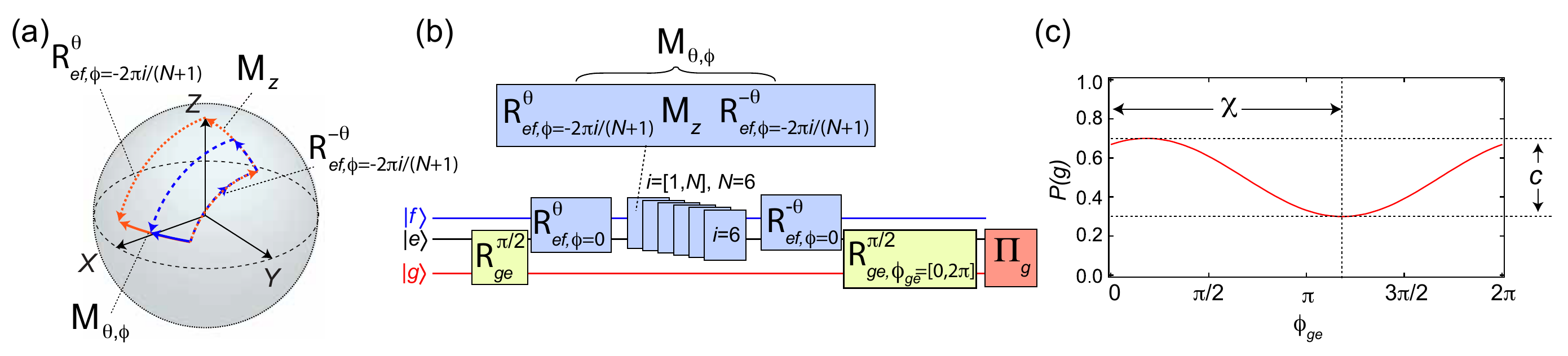}
\end{center}
\begin{flushleft}
\caption{\label{fig3} \zww{ {\bf Experimental sequence} (a) Measurements along an arbitrary axis (solid arrows) are composed of rotations before and after measurement along the $Z$ axis (dashed arrows).  Partial (projective) measurements are indicated in blue (red). (b) The full experiment consists of a sequence of 6 measurements in the qubit manifold at decreasing azimuthal angles $\phi$.  (c) The geometric phase, $\chi$, and interference \rr{contrast}, $c$, are determined from the final interference pattern between $|g\rangle$ and $|e\rangle$, \rr{where $P(g)$ is the probability of obtaining outcome $g$ from the final projective measurement, $\Pi_g$.}}}
\end{flushleft}
\end{figure*}
 
We characterize the strength and selectivity of the measurement by examining the dephasing rates of each pair of states. We drive the cavity with a weak probe, and perform Ramsey measurements on each pair of levels to determine the dephasing rates $\gamma_{ge}, \gamma_{ef}, \gamma_{gf}$, \rr{in each of the two-level subspaces}. In Figure \ref{fig2}d we display these measured dephasing rates versus probe frequency. The data show enhanced dephasing at each qutrit-state-dressed cavity resonance, as expected. We further observe larger background dephasing related to the $\ket{f}$ state which is due to the reduced charge noise insensitivity of the higher transmon levels \cite{koch07}.  A cavity probe \rs{at} frequency $\omega_\mathrm{p}/2\pi = 5.6715$ GHz therefore allows us to realize measurements on the $\{\ket{e}, \ket{f}\}$ manifold, while preserving coherence within the $\{\ket{g}, \ket{e}\}$ manifold. \rr{The measurement strength ($\gamma_{ef} \tau$) can be tuned via the duration of a single measurement.}

We now focus on the quantum dynamics of the qubit formed by the $\{\ket{e}, \ket{f}\}$ manifold, reserving the state $\ket{g}$ as a phase reference. Since the dispersive measurements merely provide measurement in the energy basis of the qubit, corresponding to the $Z$ axis of the Bloch sphere, we utilize additional rotations to  perform measurement along any arbitrary axis of the qubit \cite{camp13strob}.  One example of these rotations is shown in Figure \ref{fig3}a for cases of projective and partial measurements. 


Figure \ref{fig3}b displays the experimental sequence. To form closed path measurement-induced trajectories, we first initialize the qutrit in the state $\propto \ket{g} + \ket{\theta, \phi=0}$, where $\ket{\theta, \phi}$ specifies the qubit state in the $\{\ket{e},\ket{f}\}$ manifold in terms of  polar and azimuthal angles. We then apply a sequence of six measurements at fixed $\theta$ and with \rr{decreasing}  $\phi$  chosen to wrap the Bloch sphere.  \rr{Finally, we use rotations and a projective measurement \red{(which closes the path along the shortest geodesic \cite{sjoq00,Chruciski2004}) and determine the geometric phase.}  This \emph{final} stage of the protocol involves the following steps: i) The first rotation is applied such that the state $\ket{\theta,\phi=-2\pi}$ is rotated into $\ket{e}$. ii) We then apply a $\pi/2$ rotation to interfere $\ket{g}$  (the reference state) and $\ket{e}$ (which acquired a measurement-induced phase) followed by iii) projective measurement of $\ket{g}$ \cite{reed10}.} The resulting geometric phase, $\chi$, and \rr{interference} contrast, $c$, are determined by the phase  and amplitude of the interference in the $\{\ket{g},\ket{e}\}$ manifold (Fig.~\ref{fig3}c) \rv{\footnote{\rv{The resulting phase $\chi$ is clearly gauge-invariant as it can be observed from a physical interference process. The theoretical discussions regarding the gauge-invariance of measurement-induced geometric phase can be found in Refs. \cite{gebh19,sniz20_2}. Further, since the state $\ket{g}$ is only a spectator in the above protocol, the phase $\chi$ is the phase accumulated by $\ket{e}$.}}.}


\rs{The closed paths that acquire a specific geometric phase are associated with specific trajectories, resulting from specific sequences of measurement readouts, implying the need for postselection. This postselction is implicitly enforced via the measurement architecture that preserves the coherence \rr{in the subspace  $\{\ket{g},\ket{e}\}$}, while destroying coherences with $\ket{f}$.}
 As an \rs{example,} consider the regime of a very strong measurement where $|\mathbf{r_0}|\gg \sigma$. Here, there are effectively two measurement outcomes, $\mathbf{r}\simeq \mathbf{r_0}$ and $\mathbf{r}\simeq 0$, with backaction projecting onto $\ket{f}$ or \rs{the} $\{\ket{g}, \ket{e}\}$ manifold respectively. When the outcome is \rr{``null'',}  $\mathbf{r}\simeq 0$, the \rr{population in the}  $\{\ket{g}, \ket{e}\}$ manifold is preserved, ultimately contributing to the interference used to infer the geometric phase. \rs{In this case, there is backaction on the $\{\ket{e},\ket{f}\}$ subspace corresponding to a segment of the closed trajectory.} However, when the outcome is $\mathbf{r}\simeq \mathbf{r_0}$, the population in the $\{\ket{g}, \ket{e}\}$ manifold is eliminated, so that no contribution of the trajectory to the off-diagonal term between  $\ket{g}$ and \rr{$\{\ket{e},\ket{f}\}$} remains. Thus, only with a series of measurement outcomes $\mathbf{r}\simeq 0$ (\rr{null}-\rs{outcome} path) can the resulting geometric phase be \rs{observed}, while the other paths are naturally \rv{excluded from interference, which brings the present protocol in agreement with that of Ref. \cite{gebh19}. The postselection probability in Ref. \cite{gebh19} is mapped onto the interference contrast $c$.} \rr{As discussed in the appendix}, such selection occurs \rs{for any measurement strength}, effectively giving the \rs{null-outcome path geometric phase, without any postselection on specific sequences of measurement readouts.  }

\begin{figure}
\begin{center}
\includegraphics[width = 0.5\textwidth]{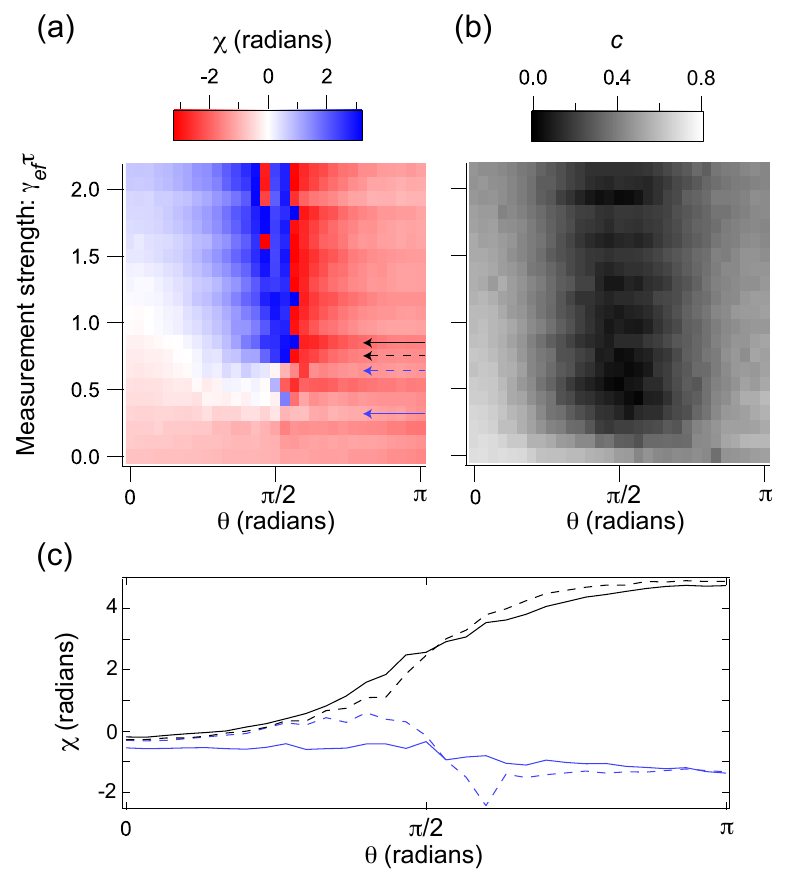}
\end{center}
\caption{\label{fig4}\zww{(a) The geometric phase under varying measurement strength ($\gamma_{ef} \tau$) gained with sequential measurements along different \rr{latitudes} of the Bloch sphere for polar angle $0\to\pi$.  \rvv{The black (blue) arrows mark measurement strengths above (below) the transition.} \rr{(b) The corresponding \rr{contrast} of the interference vanishes at the transition. \rvv{(c) The geometric phase along fixed measurement strengths before the transition (blue lines), where the set of trajectories does not wrap the Bloch sphere, and after the transition (black lines) where the set of trajectories wraps the Bloch sphere.  }  }}}
\end{figure}

In order to probe the topological transition we record the geometric phase $\chi$ and interference contrast $c$ for different trajectory latitudes and measurement \rr{strengths}. 
\rvv{The results are displayed in Fig.~\ref{fig4} and show good agreement with with the  simulation results shown in Fig.~\ref{fig1}c at measurement strengths above, below, and near the topological transition.}
In the limit of strong measurement, $\exp(-\gamma_{ef} \tau)\to 0$, the measurement backaction is sufficent to allow the quantum state to follow the measurement axis leading to \rvv{monotonically} increasing geometric phase with polar angle (as sketched in Fig.~\ref{fig1}a and c). \rr{In the infinitely weak measurement limit, \rvv{$\exp(-\gamma_{ef} \tau) \to  1$} there is no measurement backaction, hence there is no observed dependence of the geometric phase on the polar angle.} Between these two limits we encounter the topological transition, which appears as a $2\pi$ phase winding \rvv{about a singularity point}. 
\rvv{In particular, we observe that, as a function of the measurement strength, the phase along $\theta=\pi/2$ exhibits an abrupt jump of size $\pi$ \rvv{(cf. Fig.~\ref{fig4}c)} right at the singularity point, in agreement with the theory predictions (cf. Fig.~\ref{fig1}c).} 
Exactly at this point the phase is ill-defined, which can only happen if the contrast vanishes, which we indeed observe (cf. Fig.~4b). 
This \rs{jump} corresponds to the critical measurement strength that drags the state half way around the Bloch sphere \rr{(cf.\ the insets of Fig.~1c)}.  Near the transition, the state after the final projection \kwm{involves averaging over} trajectories that either encircled the Bloch sphere, acquiring a geometric phase of $\pi$, and those that did not, acquiring zero geometric phase. \rs{This leads to the observed vanishing contrast. }

We have investigated measurement enabled quantum dynamics where the dynamics carry a topological character. Despite the stochastic nature of these weak measurement dynamics, the resulting geometric phases feature a robust and sharp transition which is immune to non-universal details of the platform. \rs{A salient feature of our protocol is that right at the transition point, coherent features of the measurement-induced phase are washed out, underscoring the role of fluctuations at the critical transition point. }While our single qubit experiment has simple topology, our work \rr{advances the possibility} that many body systems with non-trivial topology might also exhibit such robust topological transitions. \rr{Such investigations would reveal the interplay between the topological nature of many-body phases and the measurement-induced dynamics. This suggests} the possibility to utilize such topological effects for protected quantum information processing.

\begin{acknowledgements}
\rs{This research was supported by NSF Grant No. PHY- 1752844 (CAREER) and used of facilities at the Institute of Materials Science and Engineering at Washington University. 
Y.G. acknowledges support by the Helmholtz International Fellow Award, the Deutsche Forschungsgemeinschaft (DFG, German Research Foundation)-- 277101999--TRR 183 (project C01) and EG 96/13-1, by the Israel Science Foundation (ISF), by the NSF Grant No. DMR-2037654, and the US-Israel Binational Science Foundation (BSF),
Jerusalem, Israel.
K.S. acknowledges funding by
Deutsche Forschungsgemeinschaft (DFG, German Research Foundation) --277101999--TRR 183 (project C01) and GO 1405/6-1.}
\end{acknowledgements}

\section*{Appendix}

\zww{\emph{Stabilizing higher Transmon states against charge noise}---Although a superconducting Transmon circuit is designed to reduce the charge noise sensitivity of the $\ket{e}$ state exponentially in the ratio between $E_\mathrm{J}/h =13.015$ GHz and $E_\mathrm{C}/h = 285$ MHz \cite{koch07}, the third energy level $|f\rangle$ may still be affected by  charge noise. We observe both increased dephasing \rr{associated with the $\ket{f}$ state} (Fig.~2c), as well as abrupt transitions in the $\ket{f}$ energy and associated fluctuations. We stabilize the experiment against these fluctuations by tracking the Ramsey pattern in the $\{ \ket{e},\ket{f}\}$ manifold and sorting the acquired data in a post-processing step.}

\begin{figure}
\begin{center}
\includegraphics[width = 0.5\textwidth]{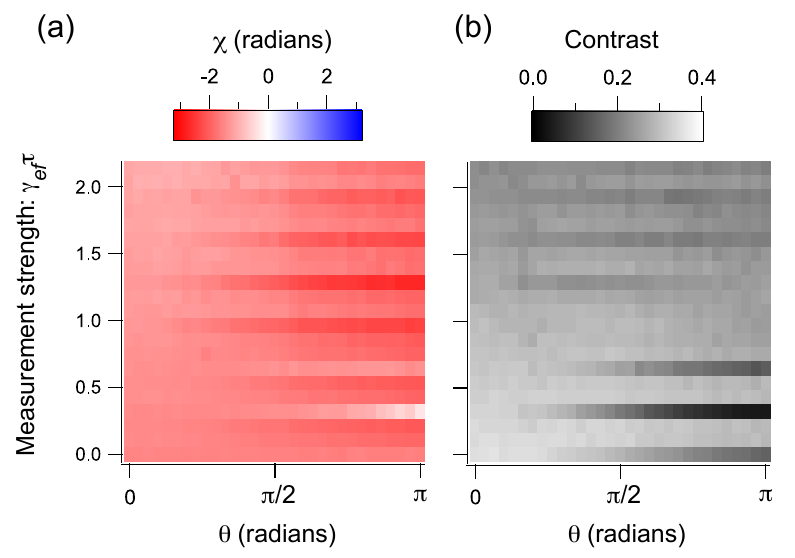}
\end{center}
\caption{\label{fig5}\rr{{\bf Cancellation of relative dynamical phases.} The observed interference phase (a) and contrast (b) for a reference experiment with fixed $\phi=0$, showing that the dynamical phase is smooth and effectively canceled through appropriate rotating frames.}}
\end{figure}

\zww{\emph{Calibrating dynamical phase accumulation}---The quantum evolution in our experiment that takes place over $\sim\mu$s of evolution is associated with \rr{the accumulation of large dynamical phases} (on the order of $10^4$ radians). This dynamical phase is measured in a rotating frame associated with microwave generator used to perform Transmon rotations. \rr{This results in an effective cancellation of the dynamical phase.} To confirm this cancellation, we perform a reference experiment at each data point using rotation sequences with fixed $\phi=0$ instead of the sequential $\phi=[0,-2\pi]$. The reference experiment makes \rr{use of} the same number and strength of sequential measurements at one point on the latitude instead of winding around the $Z$ axis. \rr{Figure \ref{fig5}  shows that this relative dynamical phase is smooth and of order a fraction of one radian, confirming that dynamical phase does not appreciably contribute to our observed topological transition. } \rr{The premise is that, while no geometric phase is accumulated, the dynamic phase gained is the same as in the actual experimental protocol.} 
\rs{The phase singularity lying at $\theta = \pi/2$ (cf. Fig. 4) further confirms that the dynamic phases do not affect the measurement-induced dynamics in our protocol, cf. Refs. 
\cite{sniz20, sniz20_2}}.
The observed stripe features in the reference phase and contrast are likely due to residual dynamical phase associated with the anharmonicity of the Transmon. }



\emph{Selective averaging theory}---The Kraus operator of the selective measurement as a function of the output signal's position on the IQ plane $\mathbf{r}$ is given by 
\begin{equation}
M_z^{(\mathbf{r})}=\left(\begin{array}{cc|c}
\tilde{\Psi}(\mathbf{r}) & 0 & 0\\
 0 & \Psi(\mathbf{r}) & 0\\
\hline
0 & 0 & \Psi(\mathbf{r})
\end{array}\right),
\end{equation}
written in the $\{\ket{f},\ket{e},\ket{g}\}$ basis, assuming unit quantum efficiency. \rr{Where $\tilde{\Psi}(\mathbf{r}) = \sqrt{\frac{1}{2 \pi}} \frac{1}{2 \sigma} e^{- (\mathbf{r}-\mathbf{r_0})^2/2\sigma^2} $ and $\Psi(\mathbf{r}) =  \sqrt{\frac{1}{2 \pi}} \frac{1}{2 \sigma} e^{- (\mathbf{r})^2/2\sigma^2} $ are as defined in the main text.}  Starting from an initial \kwm{qutrit state $\ket{\phi_{i}}=a_i^{(e)}\ket{e}+a_i^{(g)}\ket{g}$, an initial rotation $R_{0}^{\dagger}$ produces a target initial state at a chosen lattitude in the $\{\ket{e} ,\ket{f}\}$ manifold while maintaining the $\ket{g}$ state as a phase reference.  Subsequently} a sequence of measurements is performed at axes $R^\dagger_{k}\ket{e}$ with the last measurement $P_{ge}$ being projective onto the $\{\ket{g},\ket{e}\}$ manifold, leaving us with the final system state
\begin{equation}
\ket{\phi_{f}}=P_{ge}R_{0}\prod_{k}\left(R_{k}^{\dagger}M_z^{(\mathbf{r}_k)}R_{k}\right)R_{0}^{\dagger}\ket{\phi_{i}} \label{eq:phif}
\end{equation}
 for a certain series of measurement outcomes $\{\mathbf{r}_1,\mathbf{r}_2,...,\mathbf{r}_k,...\}$.

Eventually, the geometric phase $\chi$ and its \rr{contrast} $c$ \rs{are} extracted from the ensemble through the interference between state $\ket{g}$ and state $\ket{e}$, with operator $A=\Sigma_x-i\Sigma_y=2\ket{g}\bra{e}$, where
\begin{equation}
\Sigma_{x}=\left(\begin{array}{cc|c}
0 & 0 & 0\\
 0 & 0 & 1\\
\hline
0 & 1 & 0
\end{array}\right),
\\~
\Sigma_{y}=\left(\begin{array}{cc|c}
0 & 0& 0\\
0 & 0 & -i\\
\hline 
0 & i & 0
\end{array}\right),
\end{equation} 
and 
\begin{equation}
ce^{i\chi}=\int_{\Omega}\bra{\phi_f}A\ket{\phi_f}. \label{eq:contrast}
\end{equation}
Here $\Omega$ represents all the combinations for values of $\{\mathbf{r}_k\}$. \kwm{Since the operator $A$ already projects $\ket{\phi_f}$ onto the $\{\ket{g},\ket{e}\}$ manifold, we note that the final projective measurement $P_{ge}$ can be skipped in actual experiment.}

Given the initial state $\ket{\phi_i}=a_i^{(e)}\ket{e}+a_i^{(g)}\ket{g}$, \rr{after a sequence of measurements with outcomes} $\{\mathbf{r}_k\}$, the system is still in pure state $\ket{\phi_f}=a_f^{(e)}(\{\mathbf{r}_k\})\ket{e}+a_f^{(g)}(\{\mathbf{r}_k\})\ket{g}$. Since all the rotation operators $R_k$ are in the $\{\ket{e},\ket{f}\}$ manifold, the coefficient of the reference state $\ket{g}$ becomes 
\begin{equation}
a_f^{(g)}(\{\mathbf{r}_k\})=\prod_{k}\Psi(\mathbf{r}_k)a_i^{(g)}.
\end{equation} 
Using \eqref{eq:phif} and \eqref{eq:contrast}, the extracted geometric phase becomes 
$$
ce^{i\chi}=2\int_{\Omega}\bra{\phi_f}g\rangle\langle e\ket{\phi_f}=2\int_{\Omega}a_f^{(g)*}(\{\mathbf{r}_k\})\langle e\ket{\phi_f}
$$
$$
=2\bra{e}\int_{\Omega}\prod_{k}\Psi^*(\mathbf{r}_k)a_i^{(g)*}\ket{\phi_f}
$$
$$
=2a_i^{(g)*}\bra{e}\int_{\Omega}\prod_{k}\Psi^*(\mathbf{r}_k) P_{ge}R_{0}\prod_{k}\left(R_{k}^{\dagger}M_z^{(\mathbf{r}_k)}R_{k}\right)R_{0}^{\dagger}\ket{\phi_{i}}
$$
\begin{equation}
=2a_i^{(g)*}\bra{e}\int_{\Omega}R_{0}\prod_{k}\left(R_{k}^{\dagger}\Psi^*(\mathbf{r}_k)M_z^{(\mathbf{r}_k)}R_{k}\right)R_{0}^{\dagger}\ket{\phi_{i}}.
\end{equation}
Here we \kwm{note} that the whole ensemble is \kwm{a} weighted average over various measurement outcome sequences $\{r_k\}$.   \rr{For weak measurements, a null outcome corresponds to a certain probability distribution of measurement readouts $\mathbf{r}_k$. The weighting with $\Psi(\mathbf{r}_k)$ enforces the correct distribution among the states that contribute to the interference.}

Finally we have 
\begin{equation}
ce^{i\chi}=2a_i^{(g)*}\bra{e}R_{0}\prod_{k}\left(R_{k}^{\dagger} \tilde{M}_z R_{k}\right)R_{0}^{\dagger}\ket{\phi_{i}},
\end{equation}
with the integrated effective Kraus operator
$$
\tilde{M}_z=\int_{\Omega}\Psi^*(\mathbf{r})\left(\begin{array}{cc|c}
\tilde{\Psi}(\mathbf{r}) & 0 & 0\\
 0 & \Psi(\mathbf{r}) & 0\\
\hline
0 & 0 & \Psi(\mathbf{r})
\end{array}\right)
$$
$$
=\left(\begin{array}{cc|c}
\int_{\Omega}\Psi^*(\mathbf{r})\tilde{\Psi}(\mathbf{r}) & 0 & 0\\
 0 & \int_{\Omega}\Psi^*(\mathbf{r})\Psi(\mathbf{r}) & 0\\
\hline
0 & 0 & \int_{\Omega}\Psi^*(\mathbf{r})\Psi(\mathbf{r})
\end{array}\right)
$$
\begin{equation}
=\left(\begin{array}{cc|c}
\int_{\Omega}\Psi^*(\mathbf{r})\tilde{\Psi}(\mathbf{r}) & 0 & 0\\
 0 & 1& 0\\
\hline
0 & 0 & 1
\end{array}\right)=\tilde{M}_z^{\{\ket{e},\ket{f}\}}\oplus \hat{I}^{\{\ket{g}\}},
\end{equation}
where $\tilde{M}_z^{\{\ket{e},\ket{f}\}}$ is the null outcome-path partial measurement operator on the ${\{\ket{e},\ket{f}\}}$ \rv{manifold (just as the one considered in Refs. \cite{gebh19, sniz20, sniz20_2})}, and $\hat{I}^{\{\ket{g}\}}$ is the identity operator on state $\ket{g}$. 

\rr{Hence, even though the measurement readouts in our setup are continuously distributed, the effective back-action of the measurements corresponds to null-outcome partial measurements in the $\{\ket{e}, \ket{f}\}$ manifold. Thence, no explicit postselection of measurement readouts is required.}

\rr{\emph{Experimental setup}---The experiment comprises a superconducting Transmon circuit embedded in a three dimensional aluminum microwave cavity. The Transmon energies $E_\mathrm{J}/h =13.015$ GHz, $E_\mathrm{C}/h = 285$ MHz, produce transition frequencies $\omega_{ge}/2\pi = 5.12487 $ GHz, $\omega_{ef}/2\pi = 4.80788$ GHz. The cavity linewidth $\kappa/2\pi = 0.841$ MHz. The dispersive interaction between the Transmon and the cavity shifts the cavity frequency from its bare resonance frequency $\omega_\mathrm{bare}/2\pi =  5.6724$ GHz to a state dependent frequency, $\omega_g/2\pi = 5.6861$ GHz, $\omega_e/2\pi = 5.6743$ GHz, and $\omega_f/2\pi = 5.6715$ GHz.}  Three microwave generators are employed to control and measure the system, one generator addresses the Transmon transitions through single sideband modulation, another contributes the measurement at frequency $\omega_f$, and a final generator operates at $\omega_\mathrm{bare}$ to produce state readout through the Jaynes-Cummings nonlinearity technique \cite{reed10}.  The qubit/cavity is embedded in copper and magnetic shielding and cooled to a base temperature of ~10 mK in a dilution refrigerator.  The input line is subject to ~70 dB of attenuation and lossy low-pass microwave filtering. The output stage is filtered and passes through three cryogenic circulators before amplification with a HEMT amplifier.


\end{document}